\documentclass[aps,prd,reprint,superscriptaddress,nofootinbib,longbibliography]{revtex4-1}

\usepackage{amssymb,amsmath,bm,natbib}
\usepackage[dvipsnames]{xcolor}
\usepackage[colorlinks = true,
            linkcolor = Maroon,
            urlcolor  = Maroon,
            citecolor = Maroon]{hyperref}
\usepackage{microtype}
\usepackage{graphicx}
\usepackage{xspace}
\usepackage{bm}

\begin{document}

\title{An ultra-broadband axion dark matter experiment}

\author{Angelo~Esposito}
\affiliation{Dipartimento di Fisica, Sapienza Universit\`a di Roma, Piazzale Aldo Moro 2, I-00185 Rome, Italy}
\affiliation{INFN Sezione di Roma, Piazzale Aldo Moro 2, I-00185 Rome, Italy}

\author{Kin~Chung~Fong}
\affiliation{Department of Electrical and Computer Engineering, Northeastern University, Boston, MA 02115, USA}
\affiliation{Department of Physics, Northeastern University, Boston, MA 02115, USA}
\affiliation{Quantum Materials and Sensing Institute, Northeastern University, Burlington, MA}

\author{Lam~Hui}
\affiliation{Center for Theoretical Physics, Department of Physics,
Columbia University, New York, NY 10027, USA}

\date{\today}

\begin{abstract}
 	\noindent We propose a novel broadband strategy to search for axions by leveraging observables controlled by the axion field squared. We present a practical implementation of this concept for probing the axion--photon coupling.     This is done by operating a dc SQUID at the flux sweet spot, where the voltage depends quadratically on the magnetic flux, and using lock-in modulation to evade low-frequency noise. The proposed setup is ultra-broadband, spanning over 15 orders of magnitude in axion mass, with further expansion of the mass range possible.
The projected sensitivity is $|g_{a\gamma\gamma}| \gtrsim 10^{-16} \text{ GeV}^{-1}$, orders of magnitude better than current bounds, and largely independent of axion mass. We discuss the sources of systematic background and a nulling technique to reduce them to an acceptable level. We also discuss how our strategy could be adapted to probe the axion-fermion coupling, as well as to detect other dark matter candidates such as dark photons.
\end{abstract}

\maketitle


\section{Introduction}
\label{intro}

\noindent Axions-like particles (or simply axions) are pseudoscalars featured in various possible extensions of the Standard Model of particle physics~\cite[e.g.,][]{Peccei:1977hh,Weinberg:1977ma,Wilczek:1977pj,Kim:1979if,Shifman:1979if,Zhitnitsky:1980tq,Dine:1981rt,Kim:2008hd,DiLuzio:2020wdo}. They are particularly well motivated as they can offer a solution to one of the most compelling problems in fundamental physics: they could make up most of the dark matter in the Universe~\cite[e.g.,][]{Preskill:1982cy,Abbott:1982af,Dine:1982ah,Marsh:2015xka,Choi:2020rgn,Hui:2021tkt}. 
Yet, they have so far evaded all our attempts at detecting them, either directly in the lab~\cite[e.g.,][]{Irastorza:2018dyq}, or indirectly via cosmological and astrophysical observations~\cite[e.g.,][]{OHare:2024nmr,Caputo:2024oqc}.

If the mass of dark matter axions is well below $10 \text{ eV}$, their number density in the Milky Way halo is so large that the interparticle separation is smaller than the de Broglie wavelength. Their collective dynamics is then well described by that of a classical field~\cite[e.g.,][]{Sikivie:2009qn,Guth:2014hsa,Hui:2021tkt,Bao:2025nsd}, characterized by variations in time and space dictated by the axion mass and momentum, respectively. Axion dark matter experiments are usually after the signatures of such a classical field, manifesting itself as time-dependent signals in our detectors~\cite[e.g.,][]{Gramolin:2020ict,Salemi:2021gck,Adair:2022rtw,QUAX:2024fut,ADMX:2025vom,Terrano:2015sna,Wu:2019exd,Gramolin:2020ict,Arza:2021ekq,Lee:2022vvb,Nishizawa:2025xka}. However, since we currently have no hint on the value of the axion mass, these signals show up at an unknown frequency---they can potentially fall anywhere in a range covering tens of orders of magnitude. Any experiment is intrinsically limited in the range of frequencies, and hence axion masses, it can probe. This situation is even more severe for the so-called haloscopes, which achieve very large sensitivities at the cost of a very narrow frequency window. The standard solution is to carry out multiple experiments, each sensitive to different frequencies, to collectively cover the widest possible mass range.

We present a novel experimental concept which is directly sensitive to the axion field {\it squared}. Because of this, the expected signal features a zero-frequency component, independently of the axion mass. By looking for this component, one is sensitive to (almost) all axion masses, with just a single experiment. The basic idea is very simple. Consider the local axion field as a function of time:
$a(t) = a_0 {\,\rm cos} (m_at + \theta)$, where $a_0$ and $\theta$ are the amplitude and phase, respectively, and $m_a$ is the axion mass. All existing experiments involve searching for a signal that is proportional to $a$ or its derivative. Thus, the focus is on signals showing up at frequency $m_a$ (often via the power spectrum of $a$, i.e. its Fourier transform squared), and different experiments are optimized to probe a different frequency/mass range. Consider instead:
\begin{align}
    a^2(t) = {a_0^2 \over 2} + {a_0^2 \over 2} {\,\rm cos\,}\big(2 \mkern2mu m_a \mkern2mu t + 2 \mkern2mu \theta\big) \, .
\end{align}
The first term on the right is the zero-frequency component we are interested in.
It constitutes a signal we can look for, regardless of the axion mass $m_a$. 
Several issues must be addressed before realizing this idea.
(1) We need a practical way to implement the squaring. As an illustration, we will discuss how to do so by suitably configuring a Superconducting Quantum Interference Device
(SQUID). 
(2) A zero-frequency signal is often difficult to detect because of the ubiquitous random $1/f$-noise.
A standard way to get around this problem is the lock-in technique, which we will discuss.
(3)  We have essentially traded a distinctive oscillatory signal for a simple dc signal. This component is susceptible to systematic backgrounds.
We will discuss a nulling technique to reduce them to an acceptable level.
(4) A complete discussion of the expected signal ought to account for the statistical fluctuations of the axion field. As we will see, these fluctuations give rise to an additional signal, with a lineshape set by the de Broglie scale, $m_a v_a$, with $v_a$ the halo velocity dispersion.

\vspace{1em}

\noindent {\it Conventions:} Unless otherwise specified, we employ natural units $\hbar = c = \mu_0 = 1$. We also use ``frequency'' to refer to the angular frequency, $\omega$.


\section{Measurement procedure: general considerations}
\label{measurement}

\noindent How do we square the axion field, and how can we disentangle the resulting zero-frequency signal from $1/f$-noise? We consider the instance where the signal of interest is a magnetic flux proportional to the axion field, as it happens in several axion experiments~\cite[e.g.,][]{ARIADNE:2017tdd,Ouellet:2018beu,Crisosto:2019fcj,Ouellet:2019tlz,Gramolin:2020ict,Salemi:2021gck,Devlin:2021fpq,Bloch:2022kjm,DMRadio:2022pkf}.

\subsection{SQUID configuration}

\begin{figure}[t]
\centering
\includegraphics[width=\columnwidth]{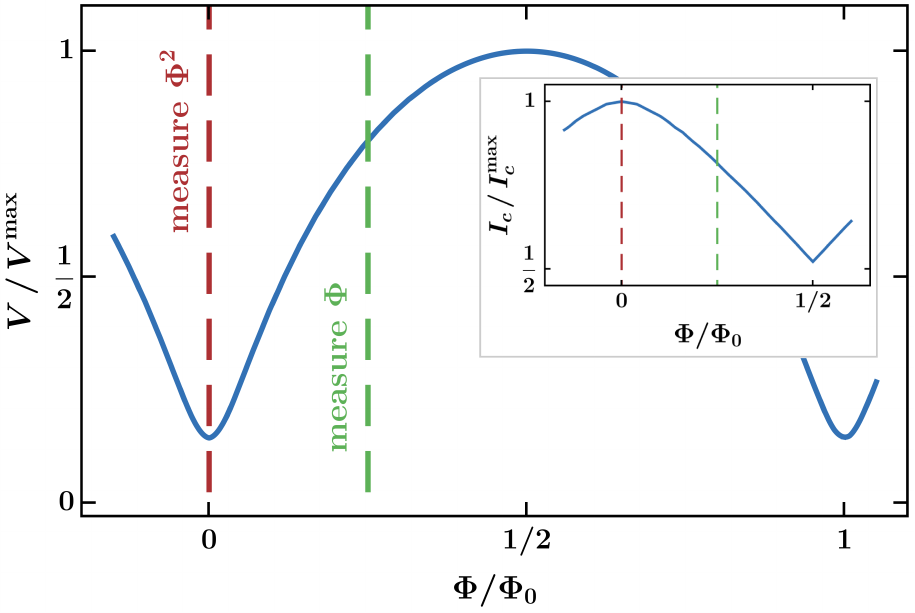} \\
\vspace{1em}
\includegraphics[width=\columnwidth]{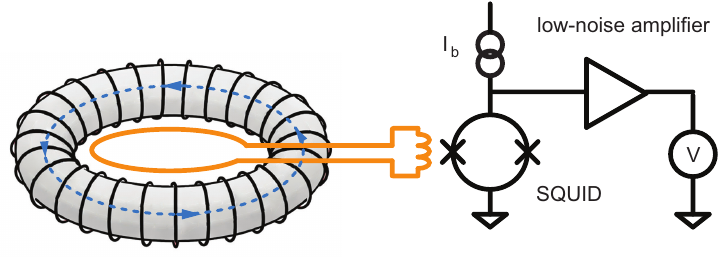}
\caption{{\bf Top panel:} Voltage response of a symmetric SQUID. When biased at the green line, the response is linear in the flux, when biased at the red line, it is quadratic in the flux. The inset shows the corresponding SQUID critical current versus flux. $\Phi_0$ is the fundamental flux quantum. {\bf Bottom panel:} Schematic representation of the experiment. As in the ABRACADABRA experiment~\cite{Ouellet:2018beu,Ouellet:2019tlz,Salemi:2021gck}, the coupling of the axion field, $a$, with the toroidal magnetic field results in an effective current. Unlike conventional SQUID magnetometry, we propose to measure $a^2$ by operating a dc SQUID at the flux sweet spot, where the $dV/d\Phi=0$.}
\label{fig:Expt Schematic}
\end{figure}

\noindent A dc SQUID is a detector sensitive to tiny changes in magnetic flux. It is composed by two Josephson junctions placed in parallel along a superconducting loop. In the presence of a suitably chosen bias current, $I_b$, a magnetic flux through the loop,  $\Phi$, gives rise to an average voltage, $V$, which can be measured.
Here, $\Phi = \Phi_b + \Delta\Phi$, where $\Phi_b$ is an applied bias flux, and $\Delta\Phi$ is the signal to be measured (proportional to $a$ or its derivative). The conventional setup is to choose $\Phi_b$ such that $d V/d\Phi$ is non-vanishing, in which case the change in voltage is linearly proportional to the signal flux, $V \propto \Delta\Phi$.
We illustrate this in Figure~\ref{fig:Expt Schematic}. The green dashed line corresponds to the conventional setup, for a {\it linear} measurement of $\Delta\Phi$.

A technical aside: the maximum bias current at zero voltage is known as the critical current, $I_c$. An illustrative plot of $I_c$ versus flux can be found in the inset of Figure~\ref{fig:Expt Schematic}.

To directly measure $\Delta\Phi^2$ instead, we can choose a value of $\Phi_b$ at which $dV/d\Phi$ vanishes. In that case, the change in voltage becomes quadratic in the signal flux:
$V \simeq \frac{1}{2} V_{\Phi\Phi} \Delta \Phi^2$, where
$V_{\Phi\Phi}$ is the second derivative evaluated at $\Phi_b$.
This corresponds to operating the SQUID magnetometer at the red dashed line
in Figure~\ref{fig:Expt Schematic}, and it is sometimes known as the ``flux sweet spot'' for transmon qubits to avoid flux noise.\footnote{Our illustration is for a symmetric SQUID, for which the desired bias flux is $\Phi_b = 0$~\cite{tesche1977dc}. For a non-symmetric SQUID, the sweet spot need not be at $\Phi_b = 0$. In practice, there is no need to be at precisely the sweet spot. One simply wants the linear term to be sub-dominant compared to the quadratic term. In fact, a small linear contribution (proportional to $a$ or its derivative) won't matter in any case, since it would not contribute to the zero-frequency signal of our primary interest.}
It is also where $d I_c / d\Phi$ vanishes (see inset). 
This way, our SQUID setup directly measures $\Delta\Phi^2$, and thus the axion field squared.\footnote{
In principle, one could construct $\Delta\Phi^2$ digitally from a data stream of
$\Delta \Phi$ as a function of time. But doing so would require sampling $\Delta\Phi$ at a high rate, especially for high axion masses.}

The parameter $V_{\Phi\Phi}$ controls how strongly the system responds to a given flux squared. We determine its value following the procedure detailed in~\cite{tesche1977dc}, by solving the SQUID equations. For a bias current that is $1\%$ larger than the junction critical current, one gets $V_{\Phi\Phi} \simeq 100 \, I_c R / \Phi_0^2$, where $I_c$ is the junction critical current, $R$ the shunt resistance, and $\Phi_0 = \pi/e$ the fundamental flux quantum. From now on, we assume the following realistic values for the critical current and shunt resistance: $I_c = 10^{-5} \text{ A}$ and $R = 10 \text{ } \Omega$. For bias currents that are even closer to the critical one, it is possible to further enhance $V_{\Phi\Phi}$, by at least an order of magnitude.

\subsection{Lock-in technique} \label{sec:lock-in}

\noindent In this detection scheme, we are after the dc signal generated by the axion field squared. However, the low-frequency region of an experiment is often plagued by the so-called $1/f$-noise, which can have multiple sources
~\cite[e.g.,][]{Dutta:1981zz,Weissman:1988zz,fagaly2006superconducting,kogan2008electronic,paladino20141}.
A standard way to circumvent this difficulty is by implementing a so-called lock-in measurement~\cite[e.g.,][]{blair1975phase,horowitz1981art,meade1983lock,zhang2024lock}. (In radio astronomy, this is referred to as ``Dicke switch''.)
It consists of two steps. First, one introduces a modulation to the signal, say, at frequency $\omega_0$. We will give a concrete method for doing so in Section~\ref{sensitivity}. If $\omega_0$ is chosen large enough, the modulation brings the interesting signal at a frequency where the $1/f$-noise is subdominant compared to other sources of noise, such as thermal or shot noise of the SQUID. Consequently, the signal can now be detected.
The second step is to multiply the detected, modulated signal by a reference signal
at the same frequency $\omega_0$ (with a suitably chosen phase).
This produces a signal that can be low-pass filtered to isolate the zero-frequency component of interest.
The key is to apply the lock-in modulation early in the measurement chain, {\it before} the sources of $1/f$-noise degrade the signal.


\section{Measurement sensitivity: an illustrative proposal}
\label{sensitivity}

\noindent We now illustrate these ideas by presenting a concrete proposal. We also discuss how to account for the stochastic fluctuations of the axion field, as well as how to take care of possible systematic backgrounds.
We are interested in probing the axion--photon coupling:
\begin{align} \label{eq:Lint}
    \mathcal{L}_{\rm int} = \frac{1}{4} \mkern2mu g_{a\gamma\gamma} \mkern2mu a \mkern2mu F_{\mu\nu} \mkern2mu \tilde F^{\mu \nu} \,,
\end{align}
where $\tilde F^{\mu\nu}$ is the dual Maxwell tensor. Further discussions concerning different couplings can be found in Section~\ref{sec:discuss}.

To connect with Section~\ref{measurement}, we need a way to generate the magnetic flux $\Delta\Phi$ from the axion.
For concreteness, we assume a configuration analogous to the one employed by the ABRACADABRA experiment~\cite{Kahn:2016aff,Ouellet:2019tlz}: an external magnetic field, $\bm{B}_0(\bm x, t)$, confined in a toroid of sizes of order $\ell$, with a circular pick-up loop at its center (see Figure \ref{fig:Expt Schematic}).
The lock-in technique is implemented by modulating $\bm{B}_0$ in time at frequency $\omega_0$.\footnote{\label{omega0} If the modulation (the ac part of $\bm{B}_0$) is generated by a coil of inductance $L \sim 1 \text{ H}$, current $I \sim 1 \text{ A}$, and sustained by a power supply providing a voltage $\Delta V \sim 1 \text{ kV}$, then the modulation frequency must be such that $L \mkern2mu dI/dt \sim L \mkern2mu \omega_0 \mkern2mu I \lesssim \Delta V$. This implies $\omega_0 \lesssim 1 \text{ kHz}$.
We envision that for an experiment constructed to minimize the $1/f$-noise,
$\omega_0$ can be kept as low as $10 - 50$~Hz.}
As shown in Appendix~\ref{app:Maxwell}, in the regime where $\omega_0 \ll 1/\ell$ and $m_a \ll 1/\ell$, the magnetic field induced by the presence of the axion, $\bm B_a(\bm x,t)$, is determined by solving the equation $\bm \nabla \times \bm B_a = - g_{a\gamma\gamma} \, \partial_t a \, \bm B_0$ in the quasistatic regime.
In this regime, the axion de Broglie wavelength is also much larger than $\ell$, $1/(m_a v_a) \gg \ell$, and the axion field can be approximated as spatially constant, $a(\bm x,t) \simeq a(t)$.
The resulting flux through the pick-up loop is,
\begin{align} \label{eq:Phia}
    \Phi_a(t) = g_{a\gamma\gamma} \, \dot a(t) \, B_{\rm in}(t) \, G \,.
\end{align}
Here $B_{\rm in}$ is the magnitude of the external magnetic field, $\bm B_0$, evaluated at the internal radius of the toroid, where it is maximal.
As mentioned, we take the external magnetic field as having both a dc and an ac component, $B_{\rm in}(t) \equiv B_{\rm dc} + B_{\rm ac} \cos(\omega_0 t)$. The factor $G$ is purely geometrical, analogous to that found in~\cite{Kahn:2016aff}. Assuming, for simplicity, a toroid of rectangular cross section, it is given by,
\begin{align}
    G \equiv \int_0^r d\bar\rho \, \bar\rho \int_{R_{\rm in}}^{R_{\rm out}} d\rho \int_0^{2\pi} d\theta \, \frac{h R_{\rm in} (\rho - \bar\rho \cos\theta)}{\tilde r^2 \sqrt{4\tilde r^2 + h^2}} \,,
\end{align}
where $r$ is the radius of the pick-up loop, $R_{\rm in}$ and $R_{\rm out}$ the internal and external radii of the toroid, $h$ its height, and $\tilde r^2 \equiv \bar \rho^2 + \rho^2 -2 \bar\rho \rho \cos\theta$. As a rough estimate, $G \sim \ell^3$.

As discussed in~\cite{Kahn:2016aff}, the flux through the SQUID, which is what we are eventually interested in, is proportional to the flux through the pick-up loop~\cite{myers2007calculated}:
\begin{align}
    \Delta \Phi \simeq \frac{\alpha}{2} \sqrt{\frac{L_\text{SQUID}}{L_\text{pick-up}}} \, \Phi_a \equiv \kappa \, \Phi_a \,.
\end{align}
Here $L_\text{SQUID}$ and $L_\text{pick-up}$ are the inductances of the SQUID and of the pick-up loop, and $\alpha$ is a parameter, typically given by $\alpha^2 \simeq 0.5$~\cite{clarke1979optimization}.
For a typical SQUID inductance of $L_\text{SQUID} \simeq 7 \text{ }\mu\text{H}$, one can achieve $\kappa \simeq 0.01$~\cite{Kahn:2016aff}. 

As explained in Section~\ref{sec:lock-in}, after the modulated flux has been converted into voltage by the SQUID, the lock-in technique multiplies the signal by $\cos(\omega_0 \mkern1mu t)$ to isolate the following term in the output voltage:
\begin{align} \label{eq:Vdemodulated}
    V(t) \to \frac{V_{\Phi\Phi}}{2} \mkern2mu \kappa^2 \mkern2mu g_{a\gamma\gamma}^2 \mkern2mu \dot a^2(t) \mkern2mu B_{\rm dc} \mkern2mu B_{\rm ac} \mkern2mu \mkern2mu G^2   \,.
\end{align}

We can now compute the voltage power spectrum in frequency space, defined via the 2-point correlator, $\langle V(\omega) V(\omega') \rangle \equiv 2\pi \mkern2mu \delta(\omega+\omega') P_V(\omega)$. Given the statistics of the axion field~\cite{Hui:2020hbq}, this turns out to be (see Appendix~\ref{app:correlators}),
\begin{align} \label{eq:PV}
    \begin{split}
        P_V(\omega) \simeq{}& V_{\Phi\Phi}^2 \mkern2mu \kappa^4 \mkern2mu g_{a\gamma\gamma}^4 \mkern2mu B_{\rm dc}^2 \mkern2mu B_{\rm ac}^2 \mkern2mu G^4 \mkern2mu \rho_a^2 \\
        & \times \bigg[ \frac{\pi}{2} \mkern2mu \delta(\omega) + 8 \frac{m_a^2}{k_a^4} \mkern2mu |\omega| \mkern2mu K_1 \!\left( \frac{4m_a |\omega|}{k_a^2} \right) \bigg] \,,
    \end{split}
\end{align}
where $K_1$ is the modified Bessel function of the second kind. Here $\rho_a \simeq 0.4 \text{ GeV/cm}^3$ is the local axion mass density~\cite{Piffl:2013mla,ou2024dark}, and $k_a \equiv \frac{2}{\sqrt{3}} m_a v_a$ its typical momentum, with $v_a \simeq 230$~km/s the velocity dispersion~\cite{Piffl:2013mla,monari2018escape}.
Here is the crucial point: the voltage power in Eq.~\eqref{eq:PV}, obtained from the magnetic flux squared, is {\it always} peaked at $\omega \simeq 0$, independently of the axion mass; this makes the strategy broadband. The only factor limiting the accessible masses comes from the magneto-quasistatic regime, which applies when $m_a \ll 1/\ell$. 
For masses comparable to or larger than $1/\ell$, our signal is quenched by the emergence of electric field effects, with the first corrections to the magneto-quasistatic limit scaling as $m_a^2 \ell^2$---see Appendix~\ref{app:Maxwell}. For a typical size $\ell \sim 1$~m, our estimate is valid as long as $m_a \ll 10^{-7}$~eV. In Section~\ref{sec:discuss} we discuss the limitations in probing higher masses, and how they might be avoided by looking for electric field signals.

The second term in Eq.~\eqref{eq:PV}, associated with $K_1$, is a lineshape
centered at zero frequency, with a width of the order of the de Broglie frequency: $m_a v_a^2$.\footnote{The expression in Eq.~\eqref{eq:PV} accounts for a Gaussian distribution of axion velocities. In detail, the de Broglie contribution will be modified by the Earth's motion through the halo.
}
For an experiment with duration $\Delta t$, this de Broglie lineshape would be distinguishable from the $\delta$-function signal for axion masses $m_a \gtrsim 2 \times 10^{-16} {\,\rm eV\,} (1 {\,\rm year}/\Delta t)$.\footnote{Eq.~\eqref{eq:PV} does not account for windowing by a finite experimental duration. The {\it windowed} power spectrum would replace the $\delta$-function by a line with a width of order $2\pi/\Delta t$. 
}
In the event of a detection of the zero-frequency signal, the natural next step would be to look for the de Broglie lineshape by probing adjacent frequencies.
For instance, for a detection bandwidth of $\delta \omega$, one could in principle look for the de Broglie line for $m_a \gtrsim 2 \times 10^{-16} {\,\rm eV\,}$ up to $3 \times 10^{-9} {\,\rm eV\,} (\delta\omega / {\,\rm Hz})$, though the de Broglie line would become more challenging to detect at higher $m_a$.\footnote{As discussed in Appendix~\ref{app:correlators}, for $2m_a \lesssim \delta\omega$, the Compton term at frequency $2m_a$ would also be the hallmark of axions.
}

To determine the sensitivity to the axion--photon coupling, we estimate the expected noise. After lock-in, this comes mostly from the SQUID white noise. The voltage thermal noise---the so-called Johnson's noise---has power spectrum given by $P_V^{\rm therm} = 4 \mkern2mu k_{\rm B} \mkern2mu T \mkern2mu R$, where $T$ is the temperature. The voltage noise coming from the current shot noise, instead, has a power spectrum given by $P_V^{\rm shot} = 2 \mkern2mu e \mkern2mu I_b \mkern2mu R^2$. At a temperature of $T = 1 \text{ mK}$, and for $I_b \simeq I_c = 10^{-5} \text{ A}$ and $R = 10 \text{ } \Omega$, we find, $P_V^{\rm therm} \simeq 5.5 \times 10^{-25} \text{ V}^2 \mkern2mu \text{s}$ and $P_V^{\rm shot} \simeq 3.2 \times 10^{-22} \text{ V}^2 \mkern2mu \text{s}$, where we reinstated SI units for future convenience. In a standard SQUID at sufficiently low temperatures, the thermal noise is subdominant to the shot noise. Nonetheless, one could engineer the dc SQUID to reduce the shot noise, whereas thermal noise is ineliminable.

To obtain a conservative estimate of our experimental sensitivity, we ignore the de Broglie contribution in Eq.~\eqref{eq:PV}, and estimate the signal-to-noise as:
\begin{align} \label{eq:SNR}
    \begin{split}
        \!\!\frac{S}{N} ={} \sqrt{\frac{\int \!\frac{d\omega}{2\pi} P_V}{P_V^{\rm noise}/\Delta t}}
        ={} \frac{ V_{\Phi\Phi} \mkern2mu \kappa^2 \mkern2mu g_{a\gamma\gamma}^2 \mkern2mu B_{\rm dc} \mkern2mu B_{\rm ac} \mkern2mu G^2 \mkern2mu \rho_a}{2\sqrt{P_V^{\rm noise}}} \mkern1mu \sqrt{\Delta t} \,,
    \end{split}
\end{align}
where $P_V^{\rm noise}$ is a generic noise power spectrum, such as those estimated above.
The expected sensitivity is obtained by requiring the signal to be larger than the noise, $S/N > 1$.
We find that our experimental scheme can be sensitive to an axion--photon coupling as small as,
\begin{align}
\label{glimit}
    |g_{a\gamma\gamma}| \gtrsim{}& 6 \times 10^{-15} \text{ GeV}^{-1} \left(\frac{0.4 \text{ GeV/cm}^3}{\rho_a}\right)^{\!\!1/2} \mkern-2mu \frac{1 \text{ m}^3}{G} \notag \\
    &{} \times \left( \frac{1 \text{ year}}{\Delta t} \right)^{\!\!1/4} \!\mkern-2mu \left(\frac{10 \text{ T}}{B_{\rm dc}}\right)^{\!\!1/2} \!\mkern-2mu \left(\frac{1 \text{ T}}{B_{\rm ac}}\right)^{\!\!1/2} \mkern-1mu \frac{0.01}{\kappa} \\
    &{} \times \left(\frac{100 \mkern2mu I_c \mkern2mu R/\Phi_0^2}{V_{\Phi\Phi}}\right)^{\!\!1/2} \!\mkern-2mu \left( \frac{P_V^{\rm noise}}{3.2 \times 10^{-22} \text{ V}^2 \mkern2mu \text{s}} \right)^{\!\!1/4} \,, \notag
\end{align}
with the shot noise power spectrum taken as the reference. 

\begin{figure}[t]
    \centering
    \includegraphics[width=\columnwidth]{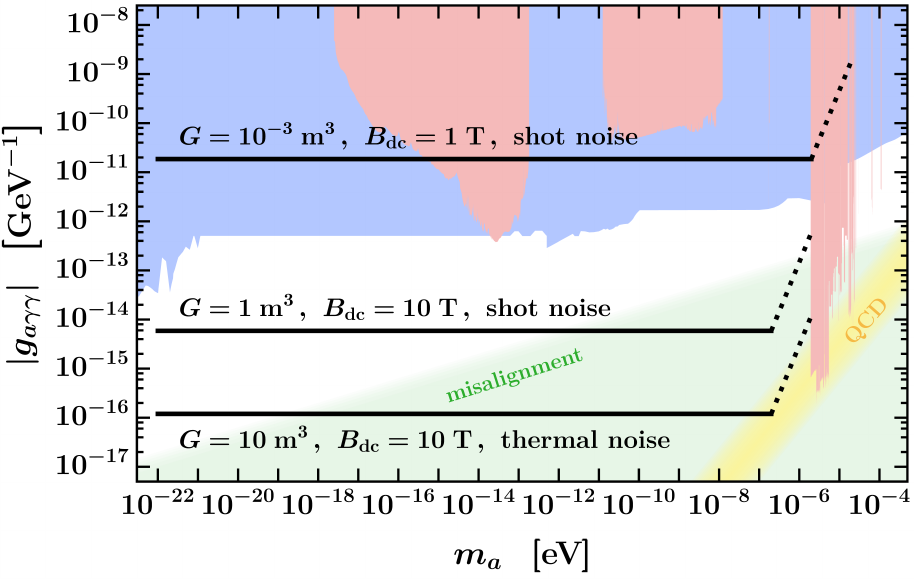}
    \caption{Projected sensitivity. Red regions are excluded by laboratory experiments, while blue regions by astrophysics and cosmology. Current bounds are taken from~\cite{AxionLimits}. The yellow band corresponds to viable QCD axion models~\cite[e.g.,][]{DiLuzio:2020wdo}, while the green region corresponds to parameters compatible with the standard misalignment mechanism~\cite[e.g.,][]{Hui:2021tkt}, assuming $|g_{a\gamma\gamma}| = C/f_a$, with $C \lesssim 1$, and a primordial misalignment angle of order unity. The sensitivity curves (black) are shown for different combinations of geometrical size of the experiment, external dc magnetic field, and noise. They all assume $\Delta t = 1 \text{ year}$, $\rho_a = 0.4 \text{ GeV/cm}^3$, $B_{\rm ac} = 1 \text{ T}$, $\kappa = 0.01$, and $V_{\Phi\Phi} = 100 \mkern2mu I_c  \mkern2mu R / \Phi_0^2$. Each curve stops at a maximum axion mass given by $1/\ell = G^{-1/3}$, as discussed in the text. The dashed lines are qualitative representations of the $m_a^2$ scaling suppressing the corrections to the magneto-quasistatic limit (Appendix~\ref{app:Maxwell}).
    For ways to probe higher masses, see Section~\ref{sec:discuss}.}
    \label{fig:sensitivity}
  \end{figure}

The comparison between the current state-of-the-art and our expected sensitivity is reported in Figure~\ref{fig:sensitivity}, for a few possible experimental configurations. 
As anticipated, the sensitivity is almost completely independent of the axion mass. A modest version of the apparatus, with a small geometrical size, and a small applied magnetic field, would already perform as well as existing experiments, but on a much broader mass range. We also stress that the existing constraints, as in Figure~\ref{fig:sensitivity}, come from about ten different laboratory experiments, and even more cosmological and astrophysical analyses. This proposal covers a wider region in parameter space with a single detector.

Our strategy of going after the zero-frequency signal from squaring raises
a natural question: how do we distinguish such a signal from
systematic backgrounds? An important source of such a background is the stray magnetic fields generated by the current coil. Ideally, the
coil should be constructed to generate a toroidal magnetic field completely
confined within it. Yet, there will be inevitable imperfections, leading to leakage. A variation of the lock-in modulation introduced earlier can also be employed to reduce such systematic backgrounds.
Details are discussed in Appendix~\ref{app:nulling}, where we show that a proper nulling technique can reduce the backgrounds to a level that wouldn't interfere with the projected sensitivity.


\section{Discussion} \label{sec:discuss}

\noindent Whether axions exist or not and, if they do, what their properties are, is one of the most compelling open questions in cosmology and particle physics. A detection scheme with the broadest possible sensitivity is then arguably highly desirable. We presented one such idea, based on the basic concept of {\it squaring} the axion field. The proposed practical implementation of this concept is to employ a suitably biased SQUID to measure the magnetic flux squared, thereby probing the axion--photon coupling in an almost completely broadband fashion.

The discussion above highlights some caveats, and offers different improvements and future directions. First, we expect some limitation when pushing the sensitivity to masses above those considered here. Indeed, when the signal frequency becomes larger, one will likely hit a number of effects limiting its observability. Examples include resonances due to the presence of any parasitic capacitance, the frequency at which the Cooper pairs of the superconducting pick-up loop fall apart, or even the time variation of the SQUID voltage. The last two are usually of order $\mathcal{O}(0.1 \text{ THz})$. Nonetheless, for masses $m_a \gg 1/\ell$ the axion-induced electric field, which is confined inside the toroid, becomes dominant. One could then envision a detector sensitive to both magnetic and electric field effects, with the latter becoming the main source of signal at large axion masses. (See also the discussion in ~\cite{Benabou:2022qpv}.)

Environmental noise showing up before the introduction of the lock-in modulation is also expected to be relevant. Besides actively minimizing it, it could also be discriminated from the signal by leveraging the properties of the latter. For example, a rotation of the pick-up loop with respect to the toroidal plane should not affect the noise, while it affects the axion signal in a predictable way. Moreover, the de Broglie lineshape is dependent on the direction of the Earth's velocity, a feature that we neglected here and that differentiates signal from noise.

Finally, the idea of squaring the signal is completely general (in fact, one might even argue, rather straightforward). One could then modify the practical implementation to make it suitable to other contexts, such as different axion couplings or even different dark matter candidates. For example, our scheme should apply equally well to the search for dark photon dark matter, which wouldn't even need the presence of the external magnetic field. By the same token, the axion--fermion coupling often manifests itself as some form of spin precession, which itself could be detected via the associated magnetic field, squared.


\begin{acknowledgments}
\noindent AE is grateful to Mauro~Valli and Bernard~van~Heck for useful discussions. KCF thanks G\"{u}n S\"{u}er for helpful discussion, and acknowledges funding support from the Department of Energy under Award Number DESC0025923. LH acknowledges support from the Department of Energy DE-SC011941, is grateful to Riccardo Penco for discussions, and thanks Sapienza University for hospitality. The views and opinions of authors expressed herein do not necessarily state or reflect
those of the United States Government or any agency thereof.
\end{acknowledgments}


\appendix

\section{Modified Maxwell's equations} \label{app:Maxwell}

\noindent In the presence of an axion field, coupled to the photon as in Eq.~\eqref{eq:Lint}, Maxwell's equations are modified as follows,
\begin{subequations} \label{eq:modifiedMaxwell}
    \begin{align}
        -\partial_t \bm E + \bm \nabla \times \bm B ={}& \bm J_0 - g_{a\gamma \gamma } \left( \partial_t a \, \bm B + \bm \nabla a \times \bm E \right) \,, \\
        \bm \nabla \cdot \bm E ={}& \rho_0 + g_{a\gamma\gamma} \bm \nabla a \cdot \bm B \,, \\
        \partial_t \bm B + \bm \nabla \times \bm E ={}& 0 \,, \\
        \bm \nabla \cdot \bm B ={}& 0 \,,
    \end{align}
\end{subequations}
where $\bm J_0$ and $\rho_0$ are the external current and charge densities, necessary to produce the external fields of interest. 

Eqs.~\eqref{eq:modifiedMaxwell} are the standard Maxwell's equations, but in the presence of a total current and charge densities which now receive contributions from the axion field as well,
\begin{subequations}
    \begin{align}
        \bm J \equiv{}& \bm J_0 - g_{a\gamma \gamma } \left( \partial_t a \, \bm B + \bm \nabla a \times \bm E \right) \,, \\
        \rho \equiv{}& \rho_0 + g_{a\gamma\gamma} \bm \nabla a \cdot \bm B \,.
    \end{align}
\end{subequations}
Taking the curl of the first equation and using the third, the magnetic field obeys a sourced wave equation with formal solution,
\begin{align}
    \bm B = \frac{1}{\partial_t^2 - \nabla^2} \bm\nabla\times\bm J \,.
\end{align}
Inverse derivatives should be intended in the operator sense, as non-local convolutions with suitable Green's functions. Now, suppose that $\bm J$ is characterized by a typical frequency $\omega$, and a typical wavelength $\ell$. If $\omega \ll \ell$, one can expand in time derivatives over gradients:
\begin{align} \label{eq:generalB}
    \bm B = - \left[ \frac{1}{\nabla^2} + \frac{\partial_t^2}{\nabla^4} + \dots \right] \bm \nabla \times \bm J \,.
\end{align}
The leading order term in this expansion is nothing but the solution to the standard Biot--Savart law~\cite[e.g.,][]{Jackson:1998nia}, indeed valid in the magneto-quasistatic limit.

At zeroth order in $g_{a\gamma\gamma}$ the equation above gives the external magnetic field, $\bm B_0$, produced by the source current, $\bm J_0$---essentially the toroidal coil. This has a typical wavelength of the order of the size of the toroid, $\ell$, and frequency given by the modulation needed for the lock-in procedure, $\omega_0$ (see Section~\ref{sec:lock-in}). For a typical size $\ell \sim 1 \text{ m}$, this means that, in order for the magneto-quasistatic equations to remain valid, the modulation frequency must be $\omega_0 \ll 1/\ell \simeq 2 \times 10^{-7} \text{ eV} \simeq 0.3 \text{ GHz}$.

Let us now turn to the order $g_{a\gamma\gamma}$ part of Eq.~\eqref{eq:generalB}. In the absence of external electric fields, the magnetic field due to the presence of the axion reads,
\begin{align} \label{eq:Bastatic}
    \bm B_a = g_{a\gamma\gamma} \left[ \frac{1}{\nabla^2} + \frac{\partial_t^2}{\nabla^4} + \dots \right] \bm \nabla  \times \left( \partial_t a \, \bm B_0 \right) \,.
\end{align}
The source on the right-hand side is now characterized by two possible frequencies: $\omega_0$, the oscillation frequency of the external magnetic field, and $m_a$, the oscillation frequency of the axion field. In order for the magneto-quasistatic expansion to remain valid, it must then also be that $m_a \ll 1/\ell \simeq 2 \times 10^{-7} \text{ eV}$. This sets a limit on the axion masses for which our approach is applicable. From Eq.~\eqref{eq:Bastatic} one further deduces that the corrections to the magneto-quasistatic limit are suppressed by $\partial_t^2/\nabla^2 \sim (m_a\ell)^2$. Since the right hand side is linear in $g_{a\gamma\gamma}$, going to higher axion masses, our sensitivity degrades as $m_a^2$.

At very large masses, the signal vanishes. Consider, in fact, the opposite regime: $m_a \gg 1/\ell$. In this case, one can instead perform an expansion in gradients over time derivatives. From Eq.~\eqref{eq:generalB} one then gets,
\begin{align} \label{eq:Blocal}
    \bm{B}_a = - g_{a\gamma\gamma} \left[ \frac{1}{\partial_t^2} + \frac{\nabla^2}{\partial_t^4} + \dots \right] \bm{\nabla} \times \left( \partial_t a \mkern2mu \bm B_0 \right) \,.
\end{align}
Since no inverse gradients appear, this solution is local in space: $\bm B_a$ vanishes where the right hand side does. Since our pick-up loop is located in a region of zero external magnetic field, the flux is also zero. Note that, given that the locality of the solution is true at all orders in the expansion~\eqref{eq:Blocal}, this implies that the signal must vanish as a non-analytical function of $(m_a \ell)^{-1}$, likely exponential.

Away from the two limiting regimes discussed above, one must solve the full wave equation~\eqref{eq:generalB} at order $g_{a\gamma\gamma}$.


\section{Axion correlators and voltage power spectrum} \label{app:correlators}

\noindent Starting from Eq.~\eqref{eq:Vdemodulated}, we can use the statistical properties of the axion field to evaluate the voltage power spectrum. Owing to the small axion velocity in the Milky Way halo, $v_a \ll 1$, we start by expanding the axion field into its non-relativistic components~\cite[e.g.,][]{Namjoo:2017nia},
\begin{align}
    a(\bm x,t) \equiv \frac{1}{\sqrt{2m_a}} \left[ e^{-im_a t} \mkern2mu \psi(\bm x,t) + e^{im_a t} \mkern2mu \psi^*(\bm x,t) \right] \,,
\end{align}
where the complex fields are such that $\partial_t \psi \ll m_a \psi$. In the present context, where the de Broglie wavelength is much larger than the typical size of the apparatus, $1/m_a v_a \gg \ell$, one can approximate the axion field as constant in space, as explained in the main text. This means to take $\psi(\bm x,t) \simeq \psi(t)$. In the infinite volume limit, the complex fields are Gaussian, and their only non-vanishing 4-point correlator is given by~\cite{Hui:2020hbq},
\begin{align}
    \begin{split}
        \left\langle \psi(t_1) \psi(t_2) \psi^*(t_3) \psi^*(t_4) \right\rangle ={}& D(t_1,t_3) D(t_2,t_4) \\
        {}&+ D(t_1,t_4) D(t_2,t_3) \,,
    \end{split}
\end{align}
where,
\begin{align}
    D(t,t') \equiv{}& \int \frac{d^3k}{(2\pi)^3} \mkern2mu \Delta(k) \mkern2mu e^{-i\omega_k(t-t')} \,.
\end{align}
The quantities appearing in the integrand are related to the properties of the local axion distribution:
\begin{align}
    \omega_k \equiv \frac{k^2}{2m_a} \,, \quad \Delta(k) \equiv \frac{\rho_a}{m_a} \mkern2mu \frac{8(2\pi)^{3/2}}{k_a^3} \mkern2mu e^{-2k^2/k_a^2} \,.
\end{align}
We recall that $k_a \equiv \frac{2}{\sqrt{3}}m_a v_a$, with $v_a$ the velocity dispersion in the Milky Way halo.

With this at hand, we determine the real-time voltage correlator in the non-relativistic limit, $\partial_t \psi \ll m_a \psi$:
\begin{widetext}
\begin{align}
    \left\langle V(t) V(t') \right\rangle \simeq{}& \left( \frac{V_{\Phi\Phi}}{2} \right)^2 \! \kappa^4 \mkern2mu g_{a\gamma\gamma}^4 \mkern2mu B_{\rm dc}^2 \mkern2mu B_{\rm ac}^2 \mkern2mu G^4 \mkern2mu \frac{m_a^2}{2} \\
    & \times \int \frac{d^3k}{(2\pi)^3} \mkern2mu \frac{d^3p} {(2\pi)^3} \mkern2mu \Delta(k) \mkern2mu \Delta(p) \bigg[ e^{-i(2m_a + \omega_k + \omega_p)(t-t')} + e^{i(2m_a+\omega_k+\omega_p)(t-t')} + 2 e^{-i(\omega_k-\omega_p)(t-t')} + 2 \bigg] \,. \notag
\end{align}
\end{widetext}
To deduce our Eq.~\eqref{eq:PV}, one moves to frequency-space,
\begin{widetext}
\begin{align}
    \left\langle V(\omega) V(\omega') \right\rangle \simeq{}& \left( \frac{V_{\Phi\Phi}}{2} \right)^2 \! \kappa^4 \mkern2mu g_{a\gamma\gamma}^4 \mkern2mu B_{\rm dc}^2 \mkern2mu B_{\rm ac}^2 \mkern2mu G^4 \mkern2mu \frac{m_a^3}{2\pi^2} \mkern2mu \delta(\omega+\omega') \notag \\
    &\times \bigg[ \Theta(|\omega|-2m_a) \mkern2mu \int_0^{\sqrt{2m_a(|\omega|-2m_a)}} dk \, k^2 \mkern2mu \sqrt{2m_a\left( |\omega| - 2m_a - \omega_k \right)} \mkern2mu \Delta(k) \mkern2mu \Delta\mkern-3mu\left(\sqrt{2m_a\left( |\omega| - 2m_a - \omega_k \right)} \mkern2mu\right) \notag \\
    & \;\quad + \frac{8\pi^4 \rho_a^2}{m_a^3} \mkern2mu \delta(\omega) + 128\pi^3 \mkern2mu \frac{\rho_a^2}{m_a k_a^4} \mkern2mu |\omega| \mkern2mu K_1\left( \frac{4m_a |\omega|}{k_a^2} \right) \bigg] \,,
\end{align}
\end{widetext}
where $\Theta(x)$ is the Heaviside step function, and $K_1$ the modified Bessel function of the second kind.
The first term in square brackets is peaked at $\omega \simeq 2m_a$. If the low-pass filter employed in the lock-in measurement selects only frequencies in a window of order $\omega_0 \simeq 50 \text{ Hz}$, this term is relevant only for axion masses $m_a \lesssim 10^{-14} \text{ eV}$. In the main text, we neglect it for simplicity. However, we note that, when visible, this term would provide a clear indication of the fact that the observed signal is indeed due to axion dark matter.


\section{Nulling the stray fluxes} \label{app:nulling}

\noindent A dangerous systematic background for the proposed scheme is expected to be leakage magnetic fields due to the imperfections of the toroidal solenoid. These generate a stray flux coming from the leakage of $B_{\rm dc}$ and $B_{\rm ac}$: $\Phi^{\rm stray}(t) = \Phi^{\rm stray}_{\rm dc} + \Phi^{\rm stray}_{\rm ac} \cos(\omega_0t)$. Once squared by the SQUID, this mimics the axion signal. 

The effect of the stray fluxes can be eliminated by a nulling procedure. This is done by introducing a ``nulling flux'' directly on the SQUID, for example via a small coil. First, one switches off the external toroidal ac field, so that $\Phi_{\rm ac}^{\rm stray} = 0$, and sets the nulling flux to $\Phi^{\rm null}(t) = \Phi_* \cos(\omega_{\rm null} t) + \Phi^{\rm null}_{\rm dc}$, with the frequency $\omega_{\rm null}$ chosen to avoid $1/f$-noise. In this configuration, the SQUID voltage features the following mode at frequency $\omega_{\rm null}$,
\begin{align}
    \begin{split}
        V(t) ={}& \frac{1}{2} V_{\Phi\Phi} \left[ \kappa \mkern2mu \Phi_a(t) + \kappa \mkern2mu \Phi^{\rm stray}(t) + \Phi^{\rm null}(t) \right]^2 \\
        \supset{}& V_{\Phi\Phi} \mkern2mu \Phi_* \left( \kappa \mkern2mu \Phi_{\rm dc}^{\rm stray} + \Phi^{\rm null}_{\rm dc} \right) \cos(\omega_{\rm null} t) \,,
    \end{split}
\end{align}
where $\Phi_a$ is the flux due to the axion field, Eq.~\eqref{eq:Phia}, evaluated at $B_{\rm ac} = 0$. At this point, one can tune $\Phi^{\rm null}_{\rm dc}$ until the mode at $\omega_{\rm null}$ disappears. When that happens, it means that $\Phi^{\rm null}_{\rm dc} \simeq -\kappa \mkern2mu \Phi_{\rm dc}^{\rm stray}$ (with an accuracy that we will specify below). To null the ac stray fluxes, one performs a similar procedure, by instead turning off the toroidal dc external field, so that $\Phi_{\rm dc}^{\rm stray} = 0$. The nulling flux is then set to $\Phi^{\rm null}(t) = \Phi_* + \Phi^{\rm null}_{\rm ac} \cos(\omega_0 t)$, such that the component of the output voltage at frequency $\omega_0$ is,
\begin{align}
    \begin{split}
        V(t) \supset V_{\Phi\Phi} \mkern2mu \Phi_* \left( \kappa \mkern2mu \Phi_{\rm ac}^{\rm stray} + \Phi^{\rm null}_{\rm ac} \right) \cos(\omega_0 t) \,.
    \end{split}
\end{align}
One then sets $\Phi^{\rm null}_{\rm ac} \simeq -\kappa \mkern2mu \Phi_{\rm ac}^{\rm stray}$ by nulling this component.
Once this is done, the dc and ac parts of the toroidal external field are both turned back on, and the nulling flux is finally set to $\Phi^{\rm null}(t) = \Phi^{\rm null}_{\rm dc} + \Phi^{\rm null}_{\rm ac} \cos(\omega_0 t)$. This way, the SQUID measures solely the axion flux, $\Delta\Phi = \kappa \mkern2mu \Phi_a(t) + \kappa \mkern2mu \Phi^{\rm stray}(t) + \Phi^{\rm null}(t) \simeq \kappa \mkern2mu \Phi_a(t)$.

For this nulling procedure to be effective, it must reduce the stray fluxes to a level smaller than the sensitivity to the axion flux. The uncertainty on the SQUID voltage is roughly $\sqrt{P_V^{\rm noise}/\Delta t}$. Thus, the stray fluxes can be compensated up to $V_{\Phi\Phi} \mkern2mu \Phi_* \left( \kappa \mkern2mu \Phi^{\rm stray} + \Phi^{\rm null} \right) \lesssim \sqrt{P_V^{\rm noise}/\Delta t}$, where we omit the ``dc'' and ``ac'' labels, as this applies to both instances.
The sensitivity to the axion flux is, instead, given by $\frac{1}{2} V_{\Phi \Phi} \left( \kappa \mkern2mu \Phi_a \right)^2 \gtrsim \sqrt{P_V^{\rm noise}/\Delta t}$.
Taking $\Phi_* \sim \Phi_0$, a measuring time $\Delta t = 1 \text{ year}$, and the shot noise power spectrum estimated in Section~\ref{sensitivity}, one finds,
\begin{align}
    \kappa \mkern2mu \Phi^{\rm stray} + \Phi^{\rm null} \lesssim 10^{-7} \times \kappa \mkern2mu \Phi_a \,,
\end{align}
making the nulling procedure effective against stray fluxes. Finally, note that a measuring time $\Delta t \gtrsim 1 \text{ year}$ allows to separate all the frequencies involved in the procedure explained above, for all relevant axion masses $m_a \gtrsim 10^{-22} \text{ eV}$.

\bibliography{biblio.bib}

\end{document}